# Band pattern formation of erythrocytes in density gradients is due to competing aggregation and net buoyancy

Felix Maurer[1,✉], Camila Romero[1], Nikolas Lerch[1], Thomas John[1], Lars Kaestner[1,2], Christian Wagner[1,3], Alexis Darras[1,4,✉]

[1]Experimental Physics, Saarland University, Campus, Saarbrücken, 66123, Saarland, Germany.
[2]Theoretical Medicine and Biosciences, Saarland University, Homburg, 66421, Saarland, Germany.
[3]Physics and Materials Science Research Unit, University of Luxembourg, Luxembourg, L-4365,Luxembourg.
[4]School of Physics, University of Bristol, Tyndall Avenue, Bristol, BS8 1TL, United Kingdom.

✉ mail@felixmilanmaurer.com ✉ alexis.charles.darras@gmail.com

Centrifugation of biological matter in density gradient solutions is a standard method for separating cell types or components. It is also used to separate red blood cells (RBCs) by age, as they lose water and become denser over their lifespan. When the density gradient is prepared with Percoll, discrete bands of RBCs are systematically observed along the gradient, despite the continuous density distribution of RBCs. Early studies suggested that cell aggregation might influence spatial distribution, but it remains debated whether a continuous density population can form discrete bands. We developed a continuity equation incorporating cell aggregation to describe the macroscopic evolution of RBC volume fraction in a density gradient, considering a continuous RBC density distribution. Numerical solutions demonstrate that the competition between isodensity distribution and aggregation is sufficient to create band patterns. Our model reproduces the temporal evolution observed in experiments, but also predicts several types of bifurcation-like behaviors for the steady-state patterns in constant gradients, depending on RBC volume fraction and aggregation energy. This demonstrates that the competition between RBC aggregation and iso-density distribution is a novel mechanism driving pattern formation.

Research into density gradient media started in the 1960s and led to advances in the separation protocols of biological matter by mass density[1–12]. Density gradients are usually prepared by concentration gradients in suspensions of nanoparticles or macromolecular solutions, with sedimentation dynamics different from that of biological objects to separate[1–5]. Among these media, suspensions of biocompatibly coated silica nanoparticles, known commercially as Percoll®, have found extensive utility in the separation of red blood cells (RBC)[5–12]. Human RBCs remain approximately 120 days in the blood circulation. During their lifetime, they experience progressive dehydration, and the mass density consequently increases[13]. Therefore, separation by density is considered a convenient way to sort RBCs according to their age[6,7,9]. Previous studies reported characteristic discontinuous RBC distributions of distinct bands, referred to as heterogeneous fractionation[5–12], which we reproduce in Fig. 1 (b,h). The underlying mechanism was unknown, sometimes attributed to intrinsic cell characteristics, e.g., heterogeneous ion exchange or cytoskeleton configurations[11]. However, early observations of RBC redistribution after a second centrifugation indicated the presence of aggregation[6]. Indeed, it is well known that, in polymer or nanoparticle suspensions, RBCs tend to aggregate into linear structures, called rouleaux, due to bridging and/or depletion forces[14–18]. Therefore, the overall aggregation energy depends on both the properties of the RBCs, as well as on the properties of the suspension[19–22]. We previously identified cell aggregation as an underlying mechanism significantly influencing band pattern formation thanks to a series of experiments where the aggregability of RBCs was altered[23]. Now, for the first time, we provide a numerical model that quantitatively predicts band pattern formation due solely to cell-cell aggregation, even if the mass density distribution is continuous.

Independently of the RBC density separation methods, the mechanics of pattern formation is a whole field of study on its own. Regular patterns, which are the result of a symmetry break, can appear in general dynamic instabilities, where two antagonistic processes compete for the dynamics of the system (e.g., Rayleigh-Bénard, Rayleigh-Taylor, and Benjamin-Feir instabilities)[24–26]. Pattern formation is also closely related to biological systems, as it determines the morphogenesis of organisms[27–32]. Paradigmatic models for microscopic systems include Turing patterns (from reaction-diffusion) and diffusion-aggregation mechanisms[33–38]. Recently, it has also been shown that Dynamic Density Functional Theory ($DDFT$) can also lead to pattern formation with repulsive interactions, or "social distancing", when modeling disease transmission in a population[39]. Among the possible patterns obtained in all these situations, band patterns are usually considered as the simplest ones, which can even appear in one-dimensional systems. However, their appearance and properties depend on the exact mechanisms and boundary conditions (e.g. as implemented in simulations, or determined by the experimental container size and shape) of the system in which they occur.

In this work, we present a continuum $DDFT$ model that describes the macroscopic evolution of the distribution of RBCs in a Percoll gradient. It demonstrates that band patterns spontaneously form because of competition between isodensity equilibrium, i.e., net buoyant forces, and intercellular aggregation forces. Therefore, we discovered the physical mechanisms leading to pattern formation in density gradients, which can reach clinical applications[12].

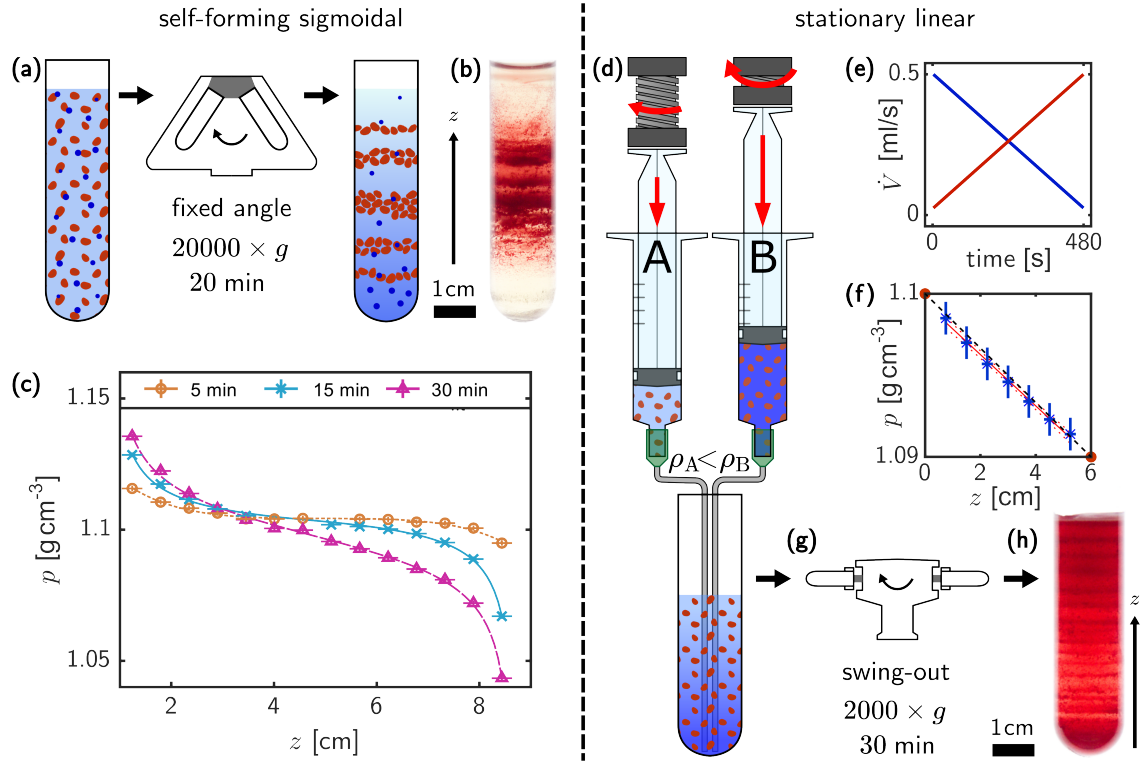

**Fig. 1 Density gradient formation and resulting band patterns**. (Left side) Self-forming gradient: **(a)** Scheme of a standard separation protocol. Packed RBCs are suspended in a Percoll medium matching the average cell density, with a hematocrit (i.e. average RBC volume fraction $\langle\psi\rangle$) of 2 %. The tube is placed in a fixed-angle rotor. Percoll particles form a concentration gradient and RBCs arrange in bands. **(b)** Photograph of the resulting experimental RBC distribution. The $z$-axis coincides with the tube axis. **(c)** Measurements of density from a self-formed gradient for different centrifugation durations $T_c$. Colored markers are measurements, while curves are regressions of a sigmoidal function (Eq.(11) in Supp. Mat.). (Right side) Stationary linear gradient **(d)** Syringe pump setup: Syringes A and B hold Percoll-based media of different densities $\rho_A < \rho_B$ and equal hematocrit $\langle\psi\rangle$. The densities are chosen symmetrically around the mean cell density. Pistons are driven by linear actuators at different speeds during 8 min in 200 steps. **(e)** The flow rate is changed linearly and reversely between A and B. **(f)** Measurements of density from a resulting linear gradient. Marker lines show error bars, red dots mark the starting densities. A linear regression (red line) is close to the expected gradient (dashed line). **(g)** The gradient is centrifuged in a swing-out rotor. **(h)** Photograph of the resulting experimental RBC distribution in the tube, displaying a band pattern, for $\langle\psi\rangle = 4$ %.

Protocols in biological or biochemical studies for age separation use spontaneous self-formation of a density gradient in Percoll, as sketched in Fig. 1**(a)**,**(b)**. Initially, a suspension of RBCs is either mixed with (or carefully placed on top of) a homogeneous Percoll medium in a centrifugation tube. The tube is then subjected to high accelerations $\geqslant 20,000 \times g$ for a duration of 20 min. Nano-sized Percoll particles then experience partial sedimentation and adopt a sigmoidal distribution. Control measurements of the density profiles without cells are shown in Fig. 1**(c)**. The density $p(z,t)$ of 14 layers extracted successively from the tube, 1 ml each, was measured using an Anton Paar Density Meter (DMA 4100M), see Fig. 1**(c)** and **(f)**. Measurements were repeated for different durations of centrifugation $T_c$. The position $z$ can empirically be linked to the Percoll density $p$ via a sigmoidal function with time-dependent slope and curvature[2–4]. The justification and properties of the chosen sigmoidal function are detailed in the Supplementary Material, and used later in the numerical model.

The RBCs suspended in the Percoll concentration gradient drift toward their isodensity position, where net buoyant force and external acceleration balance. This should result in a sorting by density. The spatial cell distribution is expected to represent the mass density distribution. However, when resolved in a Percoll gradient, the spatial distribution of RBCs organizes into a pattern of discrete and separate bands[5–12,40], see Fig. 1**(b)** and **(h)**. There is no indication that the mass density distribution of RBCs is discrete and/or multimodal. In fact, usually the distribution of the physiological RBC mass density is approximated as a Gaussian with a mean around $1.1 \, \mathrm{g\,cm^{-3}}$ and a standard deviation of $0.004 \, \mathrm{g\,cm^{-3}}$ [41,42].

Seeking a deeper explanation and mathematical description of the pattern formation process, we established a numerical model based on the assumption that the dynamics of the RBCs are governed by the acceleration of the net buoyant mass and aggregation. We first introduce the specific volume fraction $\varphi(\rho,z,t)$ of RBCs of density $\rho$ at height $z$ and time $t$, defined through its relationship to the total volume fraction $\psi(z,t) = \int d\rho \, \varphi(\rho,z,t)$. If all RBCs are supposed to have the same volume $V$, the *DDFT*-equation for this specific volume fraction $\varphi(\rho,z,t)$ can be written as

$$\zeta \partial_t \varphi = \frac{2\pi}{V} \partial_z \left( \varphi \int (z'-z)\, \tilde{u}(|z'-z|)\psi'\, \mathrm{d}z' \right) + aV \partial_z \left( \varphi\, (\rho - p(z,t)) \right), \quad (1)$$

where $\psi = \int \varphi \, d\rho$, $\zeta = 8.97 \times 10^{-8} \, \mathrm{kg\,s^{-1}}$ is the Stokes drag coefficient of an RBC with hydrodynamic radius $R = 2.8 \, \mu\mathrm{m}$, $\tilde{u}(|\mathbf{r} - \mathbf{r}'|)$ is the effective interaction potential between two RBCs located at $\mathbf{r}$ and $\mathbf{r}'$, $\mathbf{a} = -a\hat{\mathbf{z}}$ is the inertial acceleration in the centrifuge referential and $p(z,t)$ is the density of the surrounding Percoll suspension[43,44]. By this definition, $\psi(\mathbf{r}, t)$ is the volume fraction of RBCs at location $\mathbf{r}$. A common choice for the potential in case of short range interactions, like depletion, is a Lennard-Jones potential,

$$u(r) = 4U\left(\frac{\sigma^{12}}{r^{12}} - \frac{\sigma^6}{r^6}\right).$$

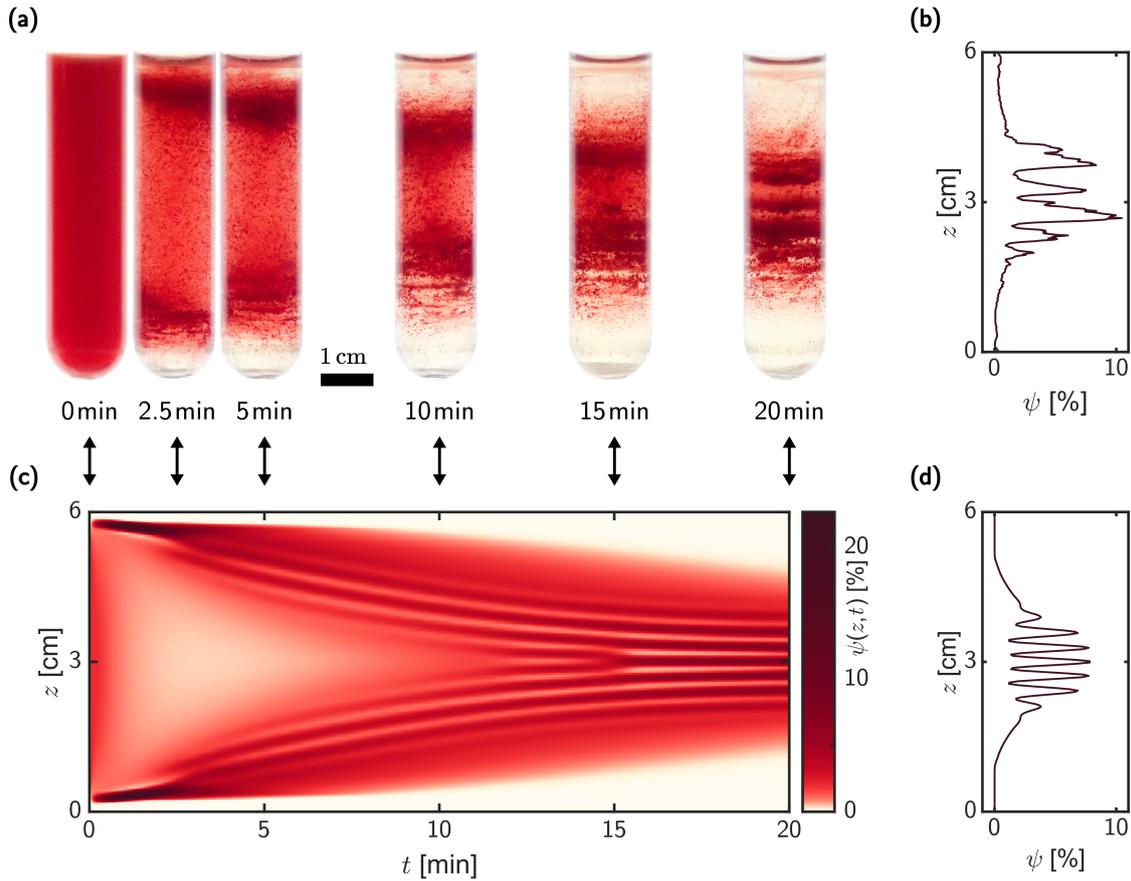

**Fig. 2** RBC fractionation in a self-forming gradient: experiment and numerical solution. **(a)** Photographs from the separation protocol after different $T_c$. The sample hematocrit was 2%. **(b)** Last experimental RBC density distribution ($t = 20$ min) showing distinct bands. $\psi(z)$ was obtained from image intensity profiles (see details in Supp. Mat.). **(c)** Numerical solution of the system with $\langle\psi\rangle = 2\%$ for comparison. **(d)** Latest numerical RBC density distribution ($t = 20$ min), showing distinct bands similar to the experiment.

The employed effective potential also takes into account the geometrical pair distribution function $g(r)$, i.e. $\tilde{u}(r) = \int g(r') \partial_r u(r') \, dr'$ [45,46]. Derivations of equation (1) and the effective potential are provided in the Supp. Mat. The numerical solver is implemented in $C^{++}$ using CUDA, see supplementary section 4 and open-access code[47]. We fix the interaction range to $\sigma = 5.6\,\mu$m, i.e. the diameter of a sphere with the volume of human RBCs, which is around 90 fL[48]. The initial condition is always homogeneous in space, which corresponds to mixing the sample tube, and a physiological distribution in density $\varphi(z, \rho, 0) = \langle\psi\rangle(2\pi\sigma_\rho^2)^{-\frac{1}{2}} e^{-(\rho - \langle\rho\rangle)^2 / 2\sigma_\rho^2}$ with $\langle\rho\rangle = 1.1\,\mathrm{g\,cm^{-3}}$, $\sigma_\rho = 4\,\mathrm{mg\,cm^{-3}}$ [41,42]. We solve the model for the traditional *self-forming sigmoidal* gradient as well as a new *stationary linear* gradient separation protocol, in order to probe the properties of the obtained pattern in a simpler case. The interaction energy is fixed and solutions for various spatial averaged volume fractions $\langle\psi\rangle$ are investigated.

The solution for the self-forming gradient protocol, illustrated in Fig. 1(a), is depicted in Fig. 2. Experiments were performed with the mean Percoll density matching the mean RBC density[41,42]. Images after different centrifugation durations are shown in Fig. 2(a). Initially, the Percoll density is constant as the medium is homogeneous, $p(z, t = 0) = p_0$. Consequently, cells with a lower density than the surrounding medium, i.e. $\rho_{\text{RBC}} < p_0$, initially move to the top and the remaining cells to the bottom. In case of a perfect density match and a symmetrical mass density distribution, this divides the population exactly in half within the first minute. The dynamics then show two different time scales related to the two different sedimenting specimens, RBCs and Percoll particles. During the first minutes, RBCs reach their isopycnic position before the gradient fully develops. The isopycnic position with respect to each cell is therefore still changing over a longer time scale until the end of centrifugation, meaning that the cell distribution follows the gradient formation dynamics. Hence, the gradient formation process also influences the pattern formation process.

The numerically calculated RBC volumetric density $\psi(z, T_c)$ is in good agreement with the experimentally observed distribution, Fig. 2(c). This includes time scales, i.e., the initial split $t < 2$ min, and the recombination with the development of a central band at around $t = 15$ min, the number and relative amplitude of bands. Differences include a break up of the distribution tails in the top and bottom of the tube due to horizontal heterogeneity, and uncontrollable variations in the experiment. Those variations might be induced by thermal or other convection, vibrations, and an offset in density matching.

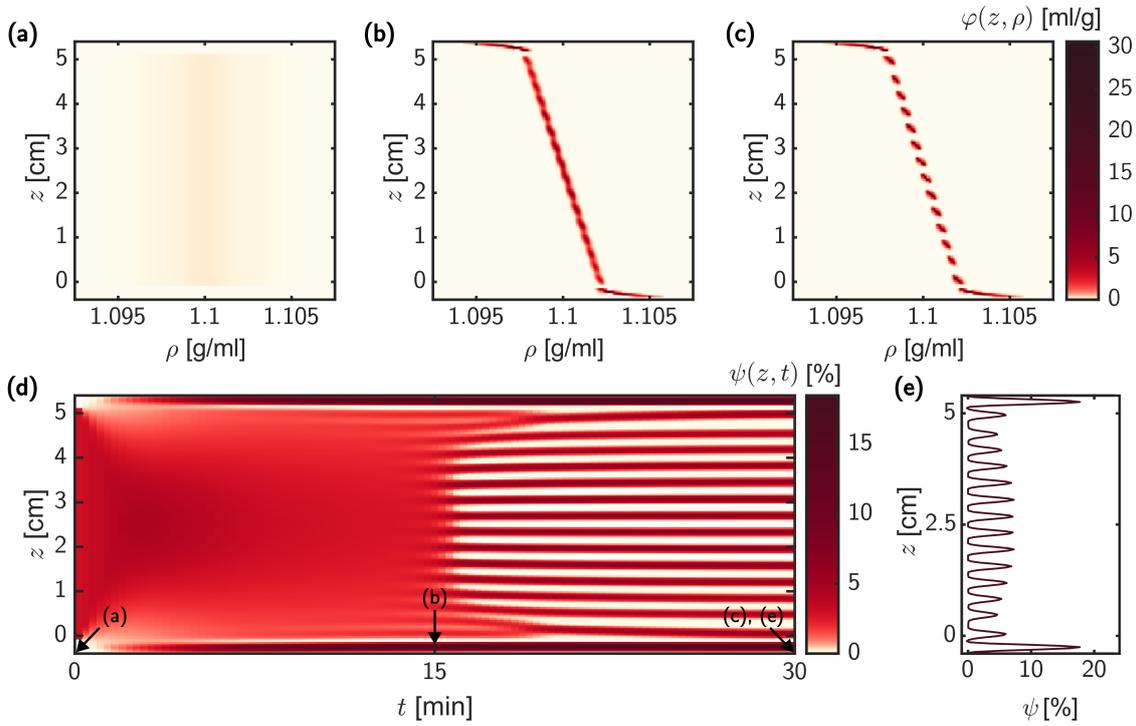

**Fig. 3** Simulation results for sedimentation in a stationary linear gradient at $\langle\psi\rangle = 3\,\%$. **(a)-(c)** RBC density function $\varphi(z,\rho,t)$ at different time points during simulation (0, 15 and 20 min, respectively). **(a)** and **(c)** show the initial homogeneous and final state, respectively, while **(b)** depicts a time point near the onset of band formation. **(d)** Time evolution of the volume fraction $\psi(z,t)$. The distribution width is smallest at around 2 min, when upward and downward moving cells are crossing around the middle of the sample. At $t \approx 15$ min distinct bands are forming which continue to sediment towards equilibrium. Also, a merging can be observed at $t \approx 17$ min. **(e)** symmetric end state showing 14 bands and tails at the borders, for which the accumulation is due to the cut-off in density within the container.

In order to perform concept-focused experiments, we designed controllable Percoll concentration gradients to study the final band pattern in various well-defined gradients and volume fractions. We use two computer-controlled syringe pumps to prepare a linear Percoll gradient and a homogeneous RBC distribution for the initial condition as shown in Fig. 1**(d)-(f)**. Syringe A holds a low density medium with the desired hematocrit $\langle\psi\rangle$, and B a high density medium, $\rho_B > \rho_A$, with the same cell concentration. The needle tips are joined in the tube bottom to achieve mixing (Fig. 1**(d)**). The volume flow is driven by linear actuators and varied to create a linear gradient (Fig. 1**(e),(f)**). Afterwards the sample is placed in a swing-out centrifuge at $2000 \times g$ for 30 min (Fig. 1**(g)**). An example of the resulting band pattern is depicted in Fig. 1**(h)**. This method uses lower accelerations, so that the distribution of Percoll does not change substantially during the sedimentation process. Hence it can be realized with swing-out rotors. It is therefore more straightforward to implement corresponding numerical simulations, as the density gradient $p(z)$ can be considered stationary along time with $\partial_z p$ constant. Fig. 3 displays a typical time evolution obtained in the corresponding numerical simulation with a stationary linear density gradient, while Fig. 4 compares experimental and numerical outcomes.

In a stationary linear gradient, a band pattern only appears above a critical concentration. Fig. 4**(a)** shows the final cell distribution after centrifugation under the same conditions for different hematocrits $\langle\psi\rangle$. It reveals a complex branching behavior in which the number of bands might increase or decrease with increasing hematocrit. Fig. 4**(c)** shows the simulation end state for different $\langle\psi\rangle$. The competition between net buoyant forces and aggregation can predict the emergence of various band patterns depending on the suspension properties and explains the high sensitivity of the band pattern on the experimental conditions. Indeed, for $\langle\psi\rangle > 3.5\,\%$, steady-state solutions exist only across a range of below $0.25\,\%$. In the experiment, the system is highly sensitive to perturbations, which can redistribute the RBCs through advection. Additionally, in 3D, we could also have heterogeneities appearing in the $xy$-plane, modifying the collision mechanisms towards a more stochastic process. Thus the observed bands are not cleanly separated, i.e. they are superimposed with a noise background. Also a symmetry breaking shift in the distribution can occur in case of a mismatch between the central gradient density and the average cell density, see Fig. 4**(a)** for $\langle\psi\rangle = 4\,\%$.

In this study we proved that, when aggregating particles with a non-unique mass density, such as RBCs, are placed in a density gradient, the assumption of an attractive pair interaction predicts band pattern formation in their spatial distribution. This occurs even if they have a monomodal and continuous mass density distribution. Our continuum model provides an explanation for the time evolution during centrifugation and final state of the RBC density function. This opens perspectives to model and predict properties of RBCs based on the observation of their band patterns, which has the potential of contributing to and enhance diagnostic predictions[12]. Moreover, preparing controlled density gradients might allow to determine conditions to obtain more sensitive or more reproducible protocols to obtain pattern formations, which can now be explained theoretically. In summary, the competition between net buoyancy and aggregation is a novel physical mechanism generating pattern formation with a complex bifurcation diagram. Its comprehension presents an exciting new challenge for researchers in the field of soft matter and biophysics, with potential clinical applications. Interestingly, although this patter formation has now been observed for the first time with biological samples, we do not need to assume any specific biological properties to observe this effect, and it should occur with artificial particles of similar sizes and geometries.

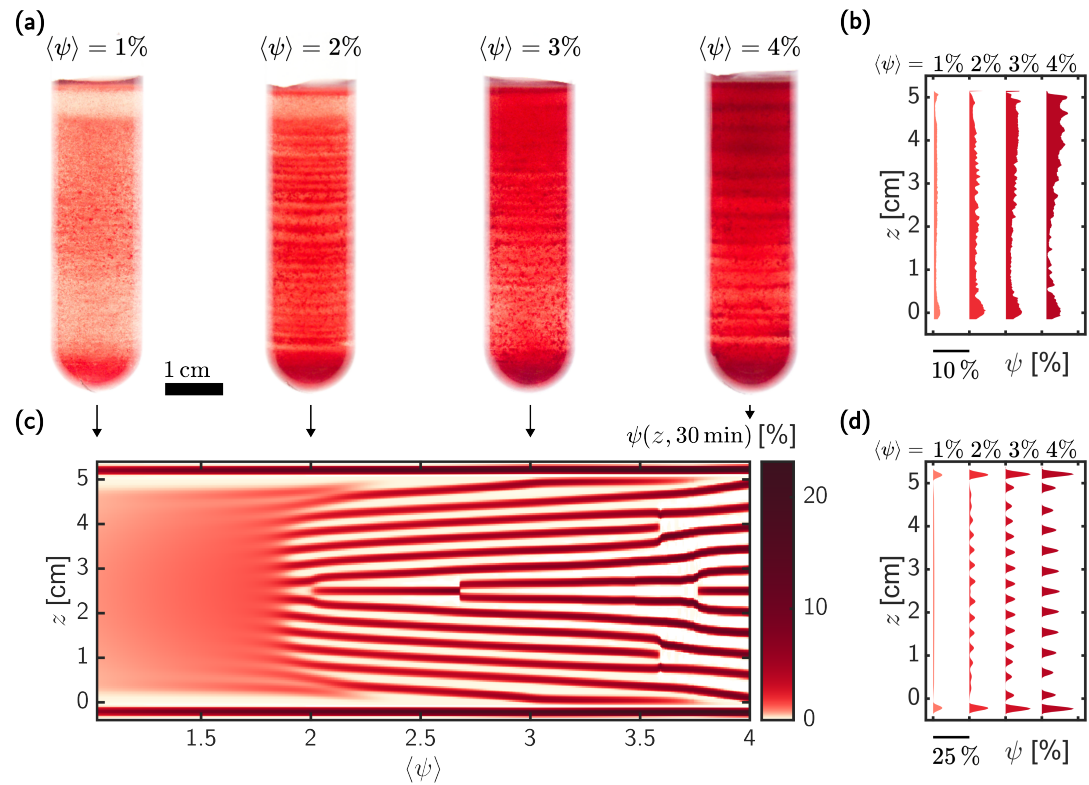

**Fig. 4** Final band structures in linear gradients, for various hematocrits $\langle\psi\rangle$. **(a)** Experimental photographs for different hematocrits. **(b)** Extracted distribution of the volume fraction, based on the image intensity profiles. Bands appear starting at hematocrits above $\approx 1\,\%$. **(c)** Simulation end state for the same system size and different volume concentrations $\langle\psi\rangle$. There is a continuous emergence of a band pattern at hematocrits between 1 and 2 %. With increasing concentration there are various bifurcations, including a change in final band positions and numbers. **(d)** Distributions of the volume fraction in numerical simulations, for selected values of $\langle\psi\rangle$.

## Author Contribution Statement

FM, TJ, LK, CW and AD developed the scientific question. FM, TJ, and AD designed the research study, and developed the initial model and solver. CW and AD obtained research funds. FM and TJ designed the experimental protocols. FM developed, performed and supervised experiments. FM wrote the final version of the model and its numerical solver, debugged, maintained and performed numerical simulations. CR investigated the time dependency of the self-forming gradient. CR and NL performed experiments. FM gathered and organized all results. FM and AD wrote the first version of the manuscript. All authors interpreted the results and wrote or reviewed the manuscript. All authors have read and agreed to the published version of the manuscript.

## Ackowledgements

A.D. acknowledges funding by the Young Investigator Grant of the Saarland University. C.W. acknowledges funding from the German Research Foundation by the research unit FOR 2688 Wa1336/12. We gratefully acknowledge the computing time granted on the supercomputer MOGON II at JGU as part of NHR South-West (nhrsw.de)[49].

## Ethical statement

Human blood withdrawal from healthy volunteers was performed after explicitly obtaining their informed consent. Blood withdrawal and handling were performed according to the declaration of Helsinki and the approval of the ethics committee "Ärztekammer des Saarlandes"(reference No 176/21).

# Band pattern formation of erythrocytes in density gradients is due to competing aggregation and net buoyancy

## Supplementary material

Felix Maurer[1,✉], Camila Romero[1], Nikolas Lerch[1], Thomas John[1], Lars Kaestner[1,2], Christian Wagner[1,3], Alexis Darras[1,4,✉]

[1]Experimental Physics, Saarland University, Campus, Saarbrücken, 66123, Saarland, Germany.
[2]Department of Theoretical Medicine and Biosciences, Saarland University, Homburg, 66421, Saarland, Germany.
[3]Physics and Materials Science Research Unit, University of Luxembourg, Luxembourg, L-4365,Luxembourg.
[4]School of Physics, University of Bristol, Tyndall Avenue, Bristol, BS8 1TL, United Kingdom.

✉ mail@felixmilanmaurer.com ✉ alexis.charles.darras@gmail.com

## 1 Establishement of the *DDFT*-equation

We will use this section of the Supplementary Materials to derive the equations exploited in the main text. To do so, we start with a general description in three spatial dimensions from which we can derive the one-dimensional description by making use of azimuthal symmetry. This reduction has implications on the shape of the interaction integral.

### 1.1 Simplified physical explanation

This section offers a simplified justification of the equations we used. Although less rigorous than the next one, it is also easier to comprehend with knowledge of classical mechanics. From a microscopic point of view, we consider the volume forces arising from the external acceleration due to centrifugation and the buoyancy following the principle of Archimedes, as well as the interaction force between cells (see Fig.S1).

The net buoyancy force combines the first two contributions. Acting on a single cell of volume $V$ and mass density $\rho$ at position $\mathbf{r}$, in a surrounding liquid of density $p(\mathbf{r},t)$, it is given by $\mathbf{f}_{\text{net}} = \mathbf{a}V(\rho - p(\mathbf{r},t))$, taking into account a constant inertial acceleration $\mathbf{a}$ due to the centrifugation of the sample. The interaction force originates from a spherically symmetric potential $\mathbf{f}_{\text{int}} = \frac{\mathbf{r}-\mathbf{r}'}{|\mathbf{r}-\mathbf{r}'|}\nabla u(|\mathbf{r}-\mathbf{r}'|)$.

Assuming that the dynamics take place in the Stokes regime, we compute cell flux terms originating in the microscopic forces. Due to conservation of the number of red blood cells, it is assumed that the volumetric cell density function $n(\mathbf{r},\rho,t)$, which describes the number of cells per volume element and mass density increment at position $\mathbf{r}$ and time $t$, will satisfy a continuity equation

$$\partial_t n(\mathbf{r},\rho,t) = -\nabla_{\mathbf{r}} \cdot \mathbf{j}(\mathbf{r},\rho,t), \quad \mathbf{j} = \mathbf{j}_{\text{net}} + \mathbf{j}_{\text{int}}. \tag{1}$$

The flux $\mathbf{j}$ associated with each force $\mathbf{f}$ is given by the product of the volumetric cell density and a characteristic cell velocity $\mathbf{j} = n\mathbf{v}$. The velocity is obtained through the ratio between the force and the cell drag coefficient $\zeta$, as $\mathbf{v} = \mathbf{f}\zeta^{-1}$. Therefore the flux associated with net buoyancy reads

$$\mathbf{j}_{\text{net}}(\mathbf{r},\rho,t) = \zeta^{-1} n(\mathbf{r},\rho,t)\,\mathbf{a} V \left(\rho - p(z,t)\right).$$

The pair-interaction between cells leads to a flux

$$\mathbf{j}_{\text{int}}(\mathbf{r},\rho,t)$$
$$= \zeta^{-1}\iint n^{(2)}(\mathbf{r},\mathbf{r}',\rho,\rho',t)\nabla_{\mathbf{r}} u(|\mathbf{r}'-\mathbf{r}|)\,\mathrm{d}^3 r'\,\mathrm{d}\rho',$$

where $u(|\mathbf{r}'-\mathbf{r}|)$ is the pair interaction potential between RBCs and $n^{(2)}(\mathbf{r},\mathbf{r}',\rho,\rho',t)$ is the two-particle density function. We use the approximation $n^{(2)}(\mathbf{r},\mathbf{r}',\rho,\rho',t) \approx g(|\mathbf{r}'-\mathbf{r}|)n(\mathbf{r},\rho,t)n(\mathbf{r}',\rho',t)$, where $g(r)$ is the *pair distribution function*[1,2].

### 1.2 General *DDFT*-equation

One can also rigorously derive the previous partial differential equation from the *DDFT*-equation

$$\frac{\partial n(\mathbf{r},\rho,t)}{\partial t} = \zeta^{-1}\nabla_{\mathbf{r}} \cdot \left(n(\mathbf{r},\rho,t)\nabla_{\mathbf{r}}\left(\frac{\delta F[n]}{\delta n(\mathbf{r},\rho,t)}\right)\right),$$

where $F[n]$ is the instrinsic Helmholtz free energy functional and $\zeta^{-1}$ is the mobility. In general there are three contributions to $F$, the energy of an ideal gas $F_{\text{id}}$, the excess energy over an ideal gas $F_{\text{exc}}$, and the contribution from an external potential $F_{\text{ext}}$[3]. The free energy of the ideal gas $F_{\text{id}}$ is irrelevant for our athermal system. The excess free energy however originates in the interactions between particles.

$$F_{\text{exc}}[n] = \frac{1}{2}\int \mathrm{d}^3 r\, \mathrm{d}^3 r'\, \mathrm{d}\rho\, \mathrm{d}\rho'\, n^{(2)}(\mathbf{r},\mathbf{r}',\rho,\rho',t)\nabla_{\mathbf{r}} u(|\mathbf{r}-\mathbf{r}'|),$$

where $u(r)$ is the effective pair interaction potential between RBCs. The external term is given by the external potential $u_{\text{ext}}(\mathbf{r},t)$,

$$F_{\text{ext}}[n(\mathbf{r},\rho,t)] = \int \mathrm{d}\rho \int \mathrm{d}^3 r\, n(\mathbf{r},\rho,t) u_{\text{ext}}(\mathbf{r},t).$$

The potential associated with the net-acceleration during sedimentation is $u_{\text{ext}}(\mathbf{r},t) = -\mathbf{r}\cdot\mathbf{a}\,V(\rho - p(\mathbf{r},t))$, with $\mathbf{a}$ the inertial acceleration due to the centrifugation process, $V$ the volume of a cell and $p(\mathbf{r},t)$ the density of the surrounding fluid (i.e. the Percoll suspension). Finally, for this system the DDFT equation then reads

$$\zeta \partial_t n =$$
$$\nabla_{\mathbf{r}} \cdot \left(n\nabla_{\mathbf{r}}\int \mathrm{d}^3 r'\mathrm{d}\rho'\, n(\mathbf{r}',\rho',t)g(|\mathbf{r}-\mathbf{r}'|)\nabla_{\mathbf{r}} u(|\mathbf{r}-\mathbf{r}'|)\right)$$
$$+ \nabla_{\mathbf{r}} \cdot (n\nabla_{\mathbf{r}}\, u_{\text{ext}}(\mathbf{r},t))$$

### 1.3 Symmetry implications and volume fraction

For our specific system, the cylindrical symmetry reduces the problem to two dimensions. Moreover, the density function is assumed to be constant in the plane orthogonal to the centrifugal acceleration $\mathbf{a}$, rendering the system one-dimensional. In cylindrical coordinates, with $\mathbf{a} = -|\mathbf{a}|\hat{\mathbf{z}}$, $|\mathbf{a}| = a$, the equation reads

$$\zeta \partial_t n$$

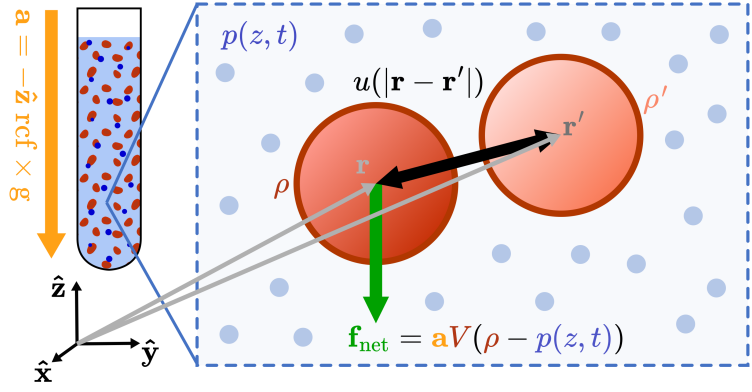

**Fig. S1 Scheme of the microscopic mechanisms involved in the numerical model.** RBCs are simplified as spheres (red) suspended in a liquid medium of density $p(z,t)$ inside of a centrifugation tube in an acceleration field. The medium contains nanoparticles (blue, not up to scale). Competing mechanisms acting on a cell are the net buoyancy force $\mathbf{f}_{\text{net}}$ and the interaction energy $u(|\mathbf{r}'-\mathbf{r}|)$ between cells with position vectors $\mathbf{r}$ and $\mathbf{r}'$.

$$= \partial_z \left( 2\pi n \int \mathrm{d}z' \mathrm{d}\rho' \, (z'-z) \, \tilde{u}(|z'-z|) n(z',\rho',t) \right)$$
$$+ aV \partial_z \left( n\left(\rho - p(z,t)\right) \right),$$

for a one-dimensional gradient $p(\mathbf{r},t) = p(z,t)$, and where $n = n(z,\rho,t)$. Here we introduce the effective interaction potential $\tilde{u}(r) = \int \mathrm{d}r' g(r') \partial'_r u(r')$. For an easier comparison with experimental results, it is practical to work with the specific volume fraction $\varphi(\mathbf{r},\rho,t) = Vn(\mathbf{r},\rho,t)$.

$$\zeta \partial_t \varphi$$
$$= \frac{2\pi}{V} \partial_z \left( \varphi \int \mathrm{d}z' \, (z'-z) \, \tilde{u}(|z'-z|) \psi' \right) \quad (2)$$
$$+ aV \partial_z \left( \varphi \left( \rho - p(z,t) \right) \right), \, \psi = \int \mathrm{d}\rho \, \varphi.$$

By that definition, $\psi(\mathbf{r},t)$ is the volume fraction of RBC at location $\mathbf{r}$. This is then the equation that we solve numerically, as explained in the main text.

## 2 Density gradient modeling

In this section the time development of the Percoll density function $p(z,t)$ is discussed. In summary, an inverse sigmoidal function, with time-dependent slope and curvature,

$$p(z,t) = p_0 + \delta(t) \frac{\chi}{\left(1 - |\chi|^{\mu(t)}\right)^{\mu(t)^{-1}}}, \; 0 \le t \le T_c, \quad (3)$$

was fitted to obtain an analytical model for the purpose of simulation[4–6]. The average density $p_0 = 1.1\,\mathrm{g\,ml^{-1}}$ in the center is assumed to be in the center of the centrifugation tube of length $L$. The density function has point symmetry with respect to the normalized coordinate $\chi = (z-z_0)\lambda^{-1}$, where $\lambda$ is a characteristic length of the gradient. There is a time-dependent density spread $\delta(t) = \delta_1 t^{\delta_2}$, and steepness $\mu(t) = \mu_1 t + \mu_2$. All coefficients $(\lambda, \delta_1, \delta_2, \mu_1, \mu_2)$ were obtained from regressions to measured densities at various durations $t = T_c$ (Fig. S2(a)). The gradient function $p(z,t)$ in Eq.(3) was adapted from regressions to experimental measurements with $\lambda = 3.38\,\mathrm{cm}$, $\delta_1 = 3.1773 \times 10^{-4}\,\mathrm{m^3 kg^{-1} s^{-\delta_2}}$, $\delta_2 = 1.52$, $\mu_1 = 1.1012 \times 10^{-3}$, and $\mu_2 = 0.6$ (Fig. S2(b)). The system length is $L = 6\,\mathrm{cm}$. The choice of this exact function is compared with the existing literature in the following paragraphs.

Maurer et al.[7] previously suggested a fit model to the measured data of centrifugation duration $T_c$,

$$p_A^{T_c}(z) = b_1 + b_2^{T_c} \log\left(\frac{b_4}{x-b_3} - 1\right), \quad (4)$$

with free parameters $b_i$. The scaling parameter $b_2$ is assumed to depend on $T_c$. After fitting the model to measurements, assumptions must be made about the time dependence of the parameter $b_2(t)$ to obtain a continuous equation for $p(z,t)$. Two variations of this model $(A)$ are compared in Fig. S3 to the model $(B)$ newly introduced for the present work. Indeed, a closer look at the measured data and different model assumptions $(A^{\mathrm{lin}}, A^{\mathrm{exp}}, B)$ justify the choice of a more elaborate mathematical equation in the numerical simulations of this article.

Laurent et al. reported a linear increase of the slope in the symmetry center of the gradient with time[5]. The first assumption $A^{\mathrm{lin}}$ is therefore a linear scaling with $t$ as given by $b_2(t) = at$, and introducing constants with a notation consistent with physical quantities, one can write

$$p(z,t) = p_0 + \log\left(\frac{\lambda}{z-z_0} - 1\right) \times at. \quad (5)$$

While the fitted curves $(A)$, displayed in Fig. S3(a), are consistent with the experimental data near the center, there is no change in curvature towards the tails. The measurements however suggest that this curvature decreases with time. For their particle model, Maurer et al.[7] described the gradient with $b_2(t) = \epsilon\left(1 - \exp(-t/\tau)\right)$,

$$p(z,t) = p_0 + \log\left(\frac{\lambda}{z-z_0} - 1\right) \times \epsilon\left(1 - \exp(-t/\tau)\right), \quad (6)$$

This assumption manages to describe the scaling of the tail, however the linear scaling near the center is lost, see Fig. S3(b).

The updated fit function in this article is given by

$$p_A^{T_c}(z) = b_1 + b_2^{T_c} \frac{\chi(z)}{\left(1 - |\chi(z)|^{b_5^{T_c}}\right)^{1/b_5^{T_c}}}, \; \chi(z) = \frac{z-b_3}{b_4}. \quad (7)$$

The fitted curves (Fig. S3(c)) correctly describe both the linear collapse near the center and the change of tail curvature. The final model, Eq. (3), then assumes the dependencies $\delta(t) = \delta_1 t^{\delta_2}$ and $\mu(t) = \mu_1 + \mu_2 t$ leading to the complete collapse shown in the article (Fig. 2(a)).

## 3 Choice of simulation parameters and pair distribution function

Centrifugation, density and viscosity measurements were consistently carried out at $22\,^\circ\mathrm{C}$. The viscosity of $80\,\%$ isotonic Percoll medium was measured at $\eta = 1.7\,\mathrm{mPa\,s}$. The depletion interaction is well documented for rigid particles, but the flexible membrane of the RBCs makes it challenging to evaluate analytically from the experimental parameters[8,9]. Although Percoll was recently identified as a strong depletion agent[7], there is currently no reference measurement to quantify the relationship between its concentration and the RBC aggregation force or energy. Also there are no measurements on the equilibrium structure of RBC aggregates in Percoll known to the authors. Due to the assumption of a short range Lennard-Jones interaction potential, we reduce the pair distribution function to the nearest neighbor. It is implemented as a gaussian peak,

$$g(|\vec{r}-\vec{r}'|) = g_0 \exp\left(-\frac{(|\vec{r}-\vec{r}'|-d)^2}{2s^2}\right),$$

with amplitude $g_0$. In ideal, $g$ would be described as a $\delta$-distribution impractical for numerical computation. The equilibrium distance is set to $d \approx 6.59\,\mu\mathrm{m}$, and we empirically choose $s = 0.5\,\mu\mathrm{m}$. An example for a potential with $U = 100\,\mathrm{aJ}$ and the corresponding effective potential is shown in Fig. S4**(a)** and **(b)**. The amplitude of the effective potential is given by $Ug_0$. For the linear gradient, we found the best match with experiments for $Ug_0 \approx 1.85 \times 10^{-9}\,\mathrm{J}$, and for the sigmoidal gradient $Ug_0 \approx 3.19 \times 10^{-9}\,\mathrm{J}$. An energy minimum of $100\,\mathrm{aJ}$ then implies $g_0 \approx 10^7$. A high nearest-neighbor amplitude indicates a regular, perhaps linear arrangement, as is the case in rouleaux formation. These assumptions are currently enough to work out an efficient approximation of the effective interaction potential. Explorations of various shapes and amplitudes of $g(r)$ will be discussed in a future publication in preparation, with comparison to experimental structures of RBC aggregates.

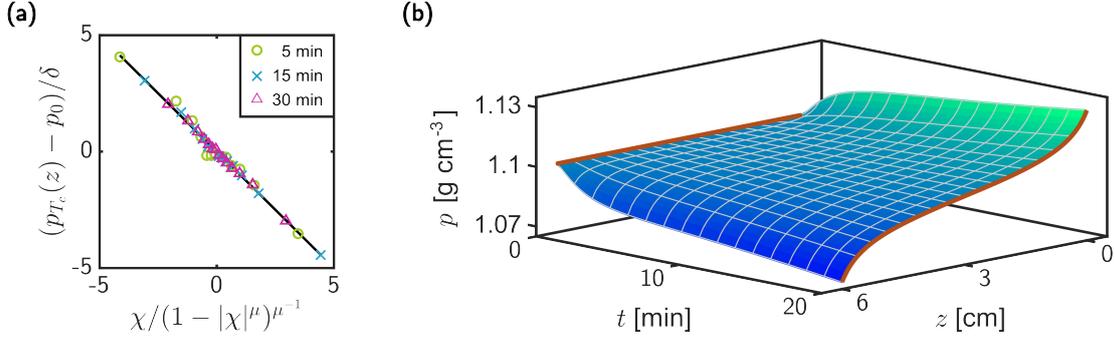

**Fig. S2** **The Percoll density function** $p(z,t)$. **(a)** Collapse of the measured densities by rescaling to linearize the fit function. The black line shows $p(z,t)$ from Eq.(3). **(b)** Surface plot of $p(z,t)$. Red lines show the density as a function of position in the beginning, $t=0$, and at $t=20\,\mathrm{min}$.

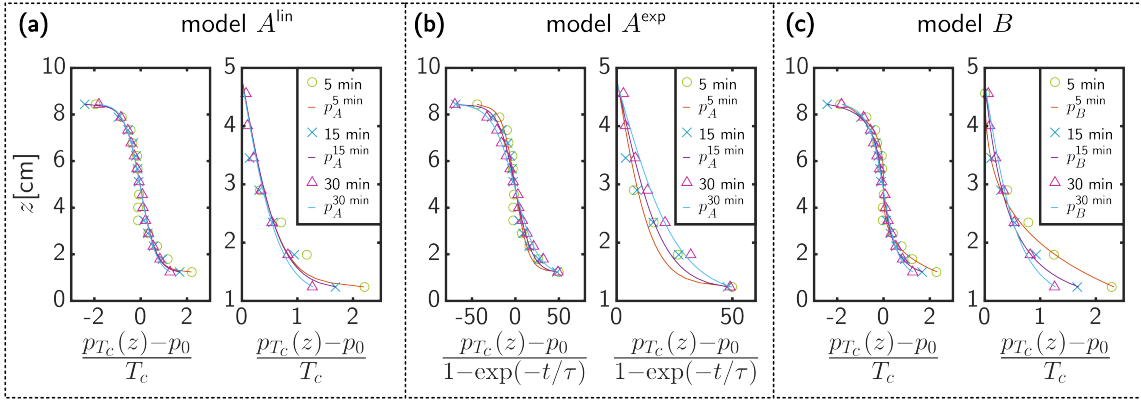

**Fig. S3** **Percoll gradient formation; comparison of different models.** **(a)** Model $A^{\mathrm{lin}}$ from Eqn. (5) assumes a linear scaling with time, but does not sufficiently reproduce the time dependent curvature. Tails of the density distribution show a divergence between experimental measurements and fitted curves. **(b)** Model $A^{\mathrm{exp}}$ from Eqn. (6), with $\tau = 20\,\mathrm{min}$, correctly fits the tails only. **(c)** Model $B$ following Eqn. (3,7) describes both the collapse near the center and captures the tails correctly.

Measurements show that the energy amplitude is comparable to the aggregation energy in autologous plasma, as it reproduces similar aggregate structures[10]. Established values show that the surface energy of the aggregation of RBCs in polymer solutions can be up to about $100\,\mu\mathrm{J/m^2}$[11].

We identify the average effective nearest-neighbor energy $\langle\tilde{u}\rangle = \frac{1}{V}\int r^2 \tilde{u}(r)\mathrm{d}r$ or rather $\epsilon = \langle\tilde{u}\rangle/U$ as the parameter with the strongest influence on the width of bands. This parameter expresses the overall balance between the attractive and repulsive interaction energy between nearest neighbors once the equilibrium, in terms of aggregate structures and $g(r)$, has been reached. Wider bands of the same amplitude require a lower value of $\epsilon$ and a higher energy $U$, see Fig. S4**(c)**. For the linear gradient it is $\langle\tilde{u}\rangle \approx -8\,\mathrm{fJ}$, for the sigmoidal gradient $\langle\tilde{u}\rangle \approx -13\,\mathrm{fJ}$, respectively.

## 4 Numerical implementation

This section provides a summary of the most important considerations of the solver implementation[12].

### 4.1 Grid and units

We work with a grid of size $N \times N = 2^k \times 2^k$. In all presented solutions it is $k = 8 \Rightarrow N = 256$. The coordinate arrays are the spatial $z_i$ and mass density $\rho_j$ discrete values. Discrete versions of density functions are $\varphi_{ij}$ and $\psi_i$. We work with $\phi_{ij} \equiv \Delta\rho\,\varphi_{ij}$, s.t. we can use the approximation $\psi_i = \sum_j \phi_{ij}$. Both quantities $\psi_i$ and $\phi_{ij}$ are dimensionless.

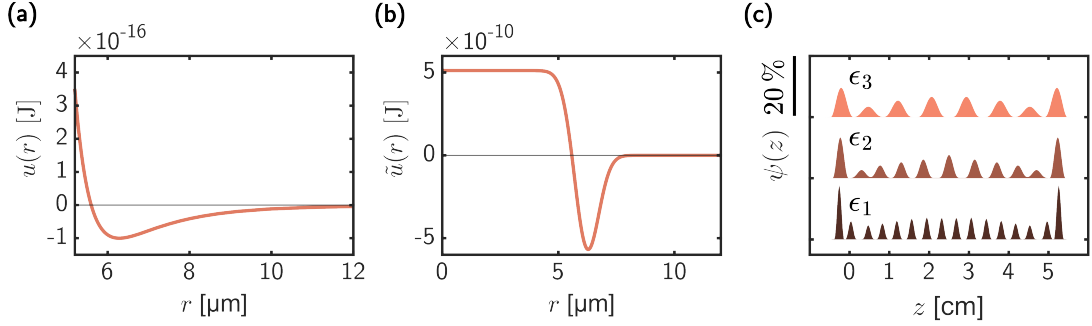

**Fig. S4** The interaction potential (a), the effective interaction potential (b) and solutions for a linear gradient and different band widths (c). The average energy was fixed to $\langle \tilde{u} \rangle \approx 7.3\,\text{fJ}$ by increasing $U$ and $\epsilon_1 \approx -73$, $\epsilon_2 \approx -32$, as well as $\epsilon_3 \approx -8.6$.

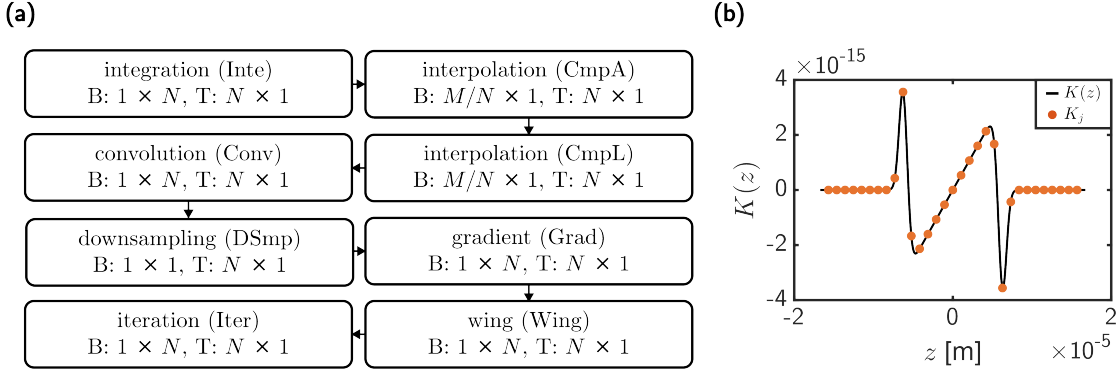

**Fig. S5** Numerical implementation. **(a)** Flow chart of CUDA kernels, brackets show abbreviations. Each kernel has its own layout in the two-dimensional grid, i.e. number of Blocks (B) and number of Threads per Block (T). **(b)** Continuous kernel function $K(z)$ and the sampled vector $K_j$ used in numerical calculations.

## 4.2 Operators

The time differential $\partial_t$ is approximated with first order forward differences with increment $\Delta t$, spatial derivatives $\partial_z$ by first order central differences, denoted by $D_z$, with increment $\Delta z$. The interaction integral is represented by a discrete convolution with the kernel array $K_j$, $j = 1, ..., 2m+1$, $m \in \mathbb{N}_0$ with an uneven number of elements. The kernel is depicted in Fig. S5(b). Since the energy is integral normalized, the sampled kernel $K_j$ has to be normalized. The deviation to the exact integral needs to be taken into account. This is implemented in form of a correction. The sampled kernel is given by $K_j = z_j \tilde{u}(z_j)$.

## 4.3 Discrete continuity equation

The continuity equation for the grid $N \times N$ reads
$$\Delta t^{-1}(\phi_{ij}^{t+1} - \phi_{ij}^t)$$
$$= \frac{aV}{\zeta} D_z \left[ \phi_{ij}^t \left( \rho_j - p_i^t \right) \right]$$
$$+ \frac{2\pi}{\zeta V} D_z \left[ \phi_{ij}^t \sum_{l=1}^{2m+1} \psi_{i+l-m-1}^t K_l \Delta z \right].$$

For performance reasons the continuity equation is solved on a coarser scale $L/N$ with the stability condition that the width of RBC bands is larger than the resolution of the coarse grid, but the kernel integration is computed on a finer grid $M$, $w_\text{band} \gg L/N > w_\text{kernel} \gg L/M$. In all the displayed solutions, it is $M = 256N = 65536$. The volume fraction $\psi_i$ is interpolated, the interaction term is calculated and finally sampled down. The upsampling is a standard cubic spline interpolation and the downsampling a simple point-wise evaluation. The numerically solved equation can then finally be written

$$\Delta t^{-1}(\phi_{ij}^{t+1} - \phi_{ij}^t)$$
$$= \frac{aV}{\zeta} D_z \left[ \phi_{ij}^t \left( \rho_j - p_i^t \right) \right] + \frac{2\pi}{\zeta V} D_z \left[ \phi_{ij}^t I_i^t \right],$$
$$I_i = \left[ \sum_{l=1}^{2m+1} \psi_{i+l-m-1}^{\text{upsmp},t} K_l \Delta z \frac{N}{M} \right]^{\text{downsmp}}.$$

## 4.4 Restrictions

The RBC volume fraction is restricted to $0 \leqslant \phi_{ij}$ and $0 \leqslant \psi_i \leqslant 1$. Those restrictions are implemented using degenerate diffusion terms[13]. Consequently, negative amplitudes due to numerical noise are prevented increasing stability of the solver. For highest stability we explicitly enforce all three conditions $\phi_{ij} \geqslant 0$, $\psi_i \geqslant 0$, and $\psi_i \leqslant 1$. The corresponding degenerate diffusion terms are $\gamma D_z^2 (1 - \phi_{ij})^{m_\text{deg}}$, $\delta D_z^2 (1 - \psi_i)^{m_\text{deg}} \phi_{ij}$, and $\kappa D_z^2 \psi_i^{m_\text{deg}} \phi_{ij}$ with an empirical value for the exponent $m_\text{deg} = 500$. The scaling factors $\gamma, \delta, \kappa \approx 10^{-11}$ are chosen to guarantee the restriction with minimal influence on the dynamics, i.e. the resulting pattern.

## 4.5 Explicit scheme

With all terms included, we get the iterative scheme for the time step $t \to t+1$.
$$\phi_{ij}^{t+1} = \phi_{ij}^t$$
$$+ \Delta t \left( \alpha D_z \left[ \phi_{ij}^t \left( \rho_j - p_i^t \right) \right] + \beta D_z \left[ \phi_{ij}^t I_i^t \right] \right)$$

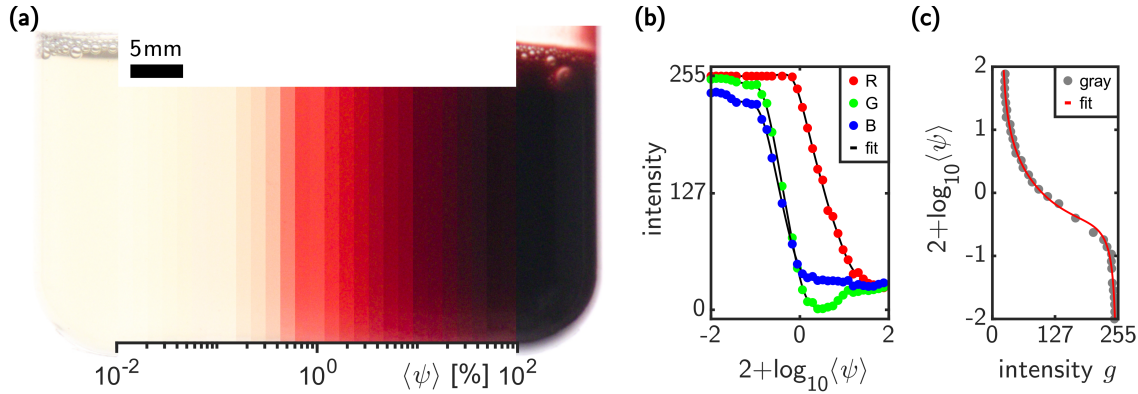

**Fig. S6 RBC concentration to RGB value conversion**. **(a)** Experimental photographs of 31 different RBC concentrations in isotonic Percoll solution. **(b)** Smoothing spline fit to each color channel. Dots mark average intensity values from a region of interest in the center of each sample tube, black lines show splines. **(c)** inverse conversion model from grayscale intensity to RBC concentration. Dots are average values of the color channels, the red line shows a regression of the model Eqn. 8.

$$-\gamma D_z^2 \left(1-\phi_{ij}\right)^{m_{\deg}} - \delta D_z^2 \left[\phi_{ij}^t \left(1-\psi_i\right)^{m_{\deg}}\right]$$
$$+ \kappa D_z^2 \left[\phi_{ij}^t \psi_i^{m_{\deg}}\right]\Big)$$

with $\alpha = \frac{aV}{\zeta}$, $\beta = \frac{2\pi}{\zeta V}$.

### 4.6 CUDA kernels

The computations during a time step are divided into different kernels, see Fig. S5**(a)**. Each kernel runs on a two-dimensional grid (blocks B : $B_x \times B_y$, threads per block T : $T_x \times T_y$) tailored to optimize efficiency. The first kernel integrates to obtain $\psi_i$. The latter is then sampled up using a standard cubic spline interpolation with natural boundary conditions. The interpolation is split into two kernels (CmpA,CmpL). The purpose of CmpA is the computation of

$$\alpha[i] = 3(\psi[i+1] - \psi[i]) - 3(\psi[i] - \psi[i-1])$$

and CmpL takes care of the iterative scheme.

$$\mu[i] = \frac{1}{4 - \mu[i-1]}$$
$$z[i] = \frac{\alpha[i] - z[i-1]}{4 - \mu[i-1]}$$
$$\mu[j] = \frac{1}{4 - \mu[j+1]}$$
$$c[j] = z[j] - \mu[j]c[j+1]$$
$$b[j] = y[j+1] - y[j] - \frac{c[j+1] + 2c[j]}{3}$$
$$d[j] = \frac{c[j+1] - c[j]}{3}$$
$$l[k] = \psi[j] + (b[j] + (c[j] + d[j]\Delta z)\Delta z)\Delta z.$$

The interpolated values are obtained directly as $\psi_k^{\text{upsmp}} = l[k]$. This is followed by the convolution kernel (Conv) and the downsampling kernel (Dsmp), that evaluates $I$ at the coarse $N$ grid points. The gradient kernel (Grad) computes the Percoll density function $p(z,t)$ for the given time step into a device array. System boundaries are implemented in form of wings extending the gradient (Wing). Finally the iteration kernel (Iter) computes flux terms, flux derivatives with first order central differences, and executes an Euler step of $\varphi_{i,j}$.

## 5 Imaging color model

This section details how RGB values from the experimental photographs have been calibrated to corresponding hematocrit values. The same color scale has been used to represent the volume fraction $\psi(z,t)$ obtained from numerical simulations, to allow for a direct visual comparison.

A total of 31 solutions of hematocrits on a logarithmic scale ranging from $10^{-2}$ to $10^2$ % were prepared. RBCs were suspended in isotonic Percoll solution to account for the interaction of light with all constituents. Each suspension was filled in the same tube used for density gradient centrifugation, and *RAW* photographs were taken under the same lighting conditions with a *Canon M50* mirrorless digital camera with the EF-M 18-150 mm zoom lens at 150 mm, aperture $f/8$, and shutter 1/60 s. Tonemapping from raw *CR3* format to RGB was performed using *RawTherapee* (rawtherapee.com), see attached *.pp3* file for detailed parameters. The image series is shown in Fig. S6**(a)**. For each photograph, a mask was drawn by hand to select a center region of each tube. The RGB channels were averaged over that area and plotted as a function of the logarithmic volume fraction in percent, $2 + \log_{10}\langle\psi\rangle$. A smoothing spline was fitted to the 8-bit intensity values $(0, ..., 255)$ of each channel using the *fit* function in *Matlab*, see Fig. S6**(b)**. The resulting regression curves were used to obtain RBC triplets from the simulated $\psi(z,t)$. The gray values $g$, i.e. the average intensity value of the color channels, were plotted versus the volume fraction $\langle\psi\rangle$ and the sigmoidal function

$$\psi^{\log}(g) = b_1 + b_2 \frac{\frac{g-b_3}{b_4}}{(1 - (\frac{g-b_3}{b_4})^{b_5})^{b_5^{-1}}} + b_6(g - b_3)^2, \quad (8)$$
$$\psi = 10^{\psi^{\log}} \%,$$

was fitted, see Fig. S6(c). The resulting parameters were $b_1 = 0.473$, $b_2 = -0.62$, $b_3 = 134.8$, $b_4 = 118.2$, $b_5 = 2.004$, $b_6 = 2.95 \times 10^{-5}$. This calibration is then later used to extract the local volume fraction $\psi(z,t)$ from the experimental pictures.